\documentclass[reprint,amsmath,amssymb,aps,pra]{revtex4-1}

\usepackage{graphicx}
\usepackage{dcolumn}
\usepackage{bm}
\usepackage{color,soul}

\begin{document}

\title{Cyber-physical risks of hacked Internet-connected vehicles}

\author{Skanda Vivek}
\thanks{These two authors contributed equally}
\affiliation{School of Physics, Georgia Institute of Technology, Atlanta, GA 30332}
                                                        
\author{David Yanni}
\thanks{These two authors contributed equally}
\affiliation{School of Physics, Georgia Institute of Technology, Atlanta, GA 30332}

\author{Peter J. Yunker}
\email{peter.yunker@gatech.edu}
\affiliation{School of Physics, Georgia Institute of Technology, Atlanta, GA 30332}

\author{Jesse L. Silverberg}
\email{js@mss.science}
\affiliation{Multiscale Systems, Inc., Division for Advanced Sciences and Data Research, Worcester MA 01609}

\date{\today}

\begin{abstract}
The integration of automotive technology with Internet-connectivity promises to both dramatically improve transportation, while simultaneously introducing the potential for new unknown risks.  Internet-connected vehicles are like digital data because they can be targeted for malicious hacking.  Unlike digital data, however, Internet-connected vehicles are cyber-physical systems that physically interact with each other and their environment.  As such, the extension of cybersecurity concerns into the cyber-physical domain introduces new possibilities for self-organized phenomena in traffic flow.  Here, we study a scenario envisioned by cybersecurity experts leading to a large number of Internet-connected vehicles being suddenly and simultaneously disabled.  We investigate post-hack traffic using agent-based simulations, and discover the critical relevance of percolation for probabilistically predicting the outcomes on a multi-lane road in the immediate aftermath of a vehicle-targeted cyber attack.  We develop an analytic percolation-based model to rapidly assess road conditions given the density of disabled vehicles and apply it to study the street network of Manhattan (NY, USA) revealing the city's vulnerability to this particular cyber-physical attack.  
\end{abstract}

\maketitle

\section*{\label{sec:Intro}Introduction}

\begin{figure}
\includegraphics[width=\columnwidth]{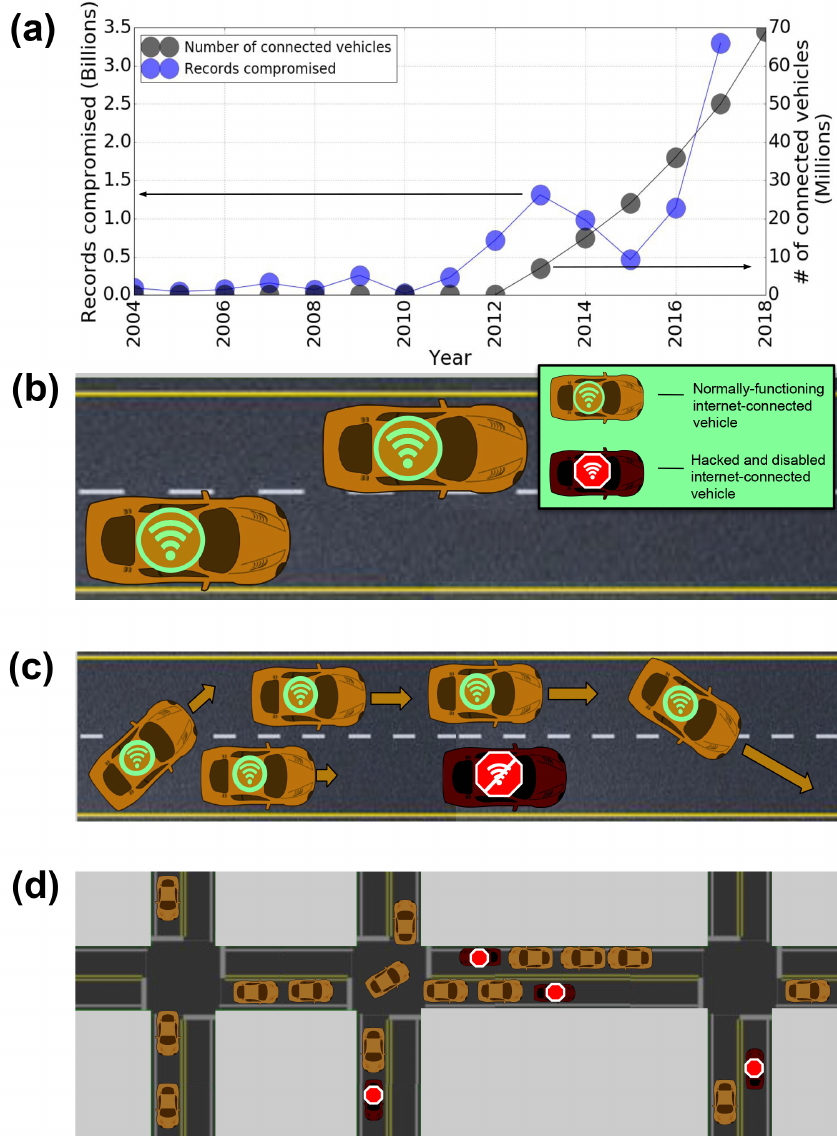}
\caption{Potential cyber-physical disruption from hacking of Internet-connected vehicles.  (\textbf{a}) Historical annual data for total number of Internet-connected vehicles (gray) and total number of digital records compromised by hacking (blue).  (\textbf{b}) Schematic of two Internet-connected vehicles traveling unobstructed on a straight 2-lane road.  (\textbf{c}) Schematic of traffic flow when an Internet-connected vehicle is disabled (red) and other vehicles must navigate around the obstacle.  (\textbf{d}) Schematic illustrating how multiple simultaneously disabled vehicles disrupts traffic flow on a network of roads.}
\label{fig:fig1}
\end{figure}

In the United States, over 40 million Internet-connected vehicles are on the road today, with hundreds of millions more expected by 2023 (Fig.~\ref{fig:fig1}a, gray circles)~\cite{statistica}.  Connected vehicle technologies have the potential to transform transportation by preventing accidents, reducing congestion, and even improving in-vehicle worker productivity~\cite{DOT-benefits}. However, these highly anticipated benefits come with largely unknown risks, especially since connected vehicles are potential targets for computer hacking~\cite{miller2015,greenberg2015,parkinson2017,amoozadeh2015}. In light of the growing number of hacking incidents exposing personal data (Fig.~\ref{fig:fig1}a, blue circles)~\cite{hacks}, cybersecurity experts are working to preemptively resolve similar software vulnerabilities and keep Internet-connected vehicles secure from similarly malicious activity~\cite{GAO-DOT,parkinson2017,eiza2017}. Nevertheless, in the event of a successful hack, compromised vehicles carry unknown cyber-physical risks, making it difficult to assess the mode and scale of disruption presented by this increasingly plausible scenario~\cite{MIT}.

While a full accounting of the risks presented by Internet-connected vehicles remains elusive, this open-ended question hasn't stopped active exploration of the possibilities.  Already, a number of scenarios have been outlined to identify how hacking increases the risk of collisions~\cite{GAO-DOT,amoozadeh2015,axelrod2017} and traffic disruptions~\cite{axelrod2017}.  For example: (i) degraded sensor input or distorted control protocols could cause connected-vehicle collisions~\cite{amoozadeh2015}, (ii) self-monitoring anti-virus-like software could cause compromised vehicles to enter a ``safe mode'' when problems are detected and bring the vehicle to a stop, reducing the likelihood of accidents but increasing localized traffic congestion~\cite{parkinson2017,Cal2018-article,Cal2018-rules}, (iii) similar self-monitoring software could directly lead to human-failure if control of a compromised vehicle was returned to an unprepared or distracted driver~\cite{richtel2015google}, and (iv) hacked sensors could be used to falsely report traffic or other objects on the road, thus inappropriately halting motion of compromised vehicles~\cite{parkinson2017,bissmeyer2010intrusion}.  Furthermore, as driver-assisting and ``auto-pilot'' technologies continue to be incorporated in modern vehicles, we can anticipate an even deeper integration between mechanical components and software-controlled systems.  In fact, some manufacturers already use wireless over-the-air updates to regularly upgrade and maintain their vehicle's software.  Thus, given the increased exposure to potential cybersecurity vulnerabilities, the growing physical control by these cybersystems over vehicular motion, and the historical precedent that ``if it can be hacked, it will be hacked,'' we can foresee the general contours of an emerging threat.  In particular, many of the cybersecurity scenarios being considered lead to a common outcome where compromised vehicles cease motion and effectively become traffic-disrupting obstacles (Fig.~\ref{fig:fig1}b and c).  From the physical perspective, we can ask what this outcome means for transportation as we quantify the emergent consequences of this cyber-physical risk. 

The scenario we investigate here is one in which a substantial number of vehicles are simultaneously disabled in a single event causing them to become immobile obstacles on the road (Fig.~\ref{fig:fig1}d).  By setting aside specifics for how cybersecurity vulnerabilities are exploited, we can instead focus on the general outcomes.  To this end, we simulate traffic flow before and after a hack has occurred so that non-compromised vehicles continue to navigate around compromised vehicles wherever possible.  Surprisingly, we find a relatively modest density of compromised vehicles can immediately halt all traffic flow.  By deriving an analytical model based on percolation theory, we show the underlying cause of this result is the local geometric arrangement of vehicles.  Application of our model to Manhattan (New York City, New York, USA) reveals the threshold number of compromised vehicles that causes city-wide gridlock and quantifies how access to emergency services is reduced.  These insights provided by this model suggest how the risks of large-scale hacks can be addressed via network compartmentalization and redundancy.

\section*{\label{sec:Results}Results and Discussion}

\subsection*{Compromised vehicles impair traffic}

\begin{figure}
\includegraphics[width=\columnwidth]{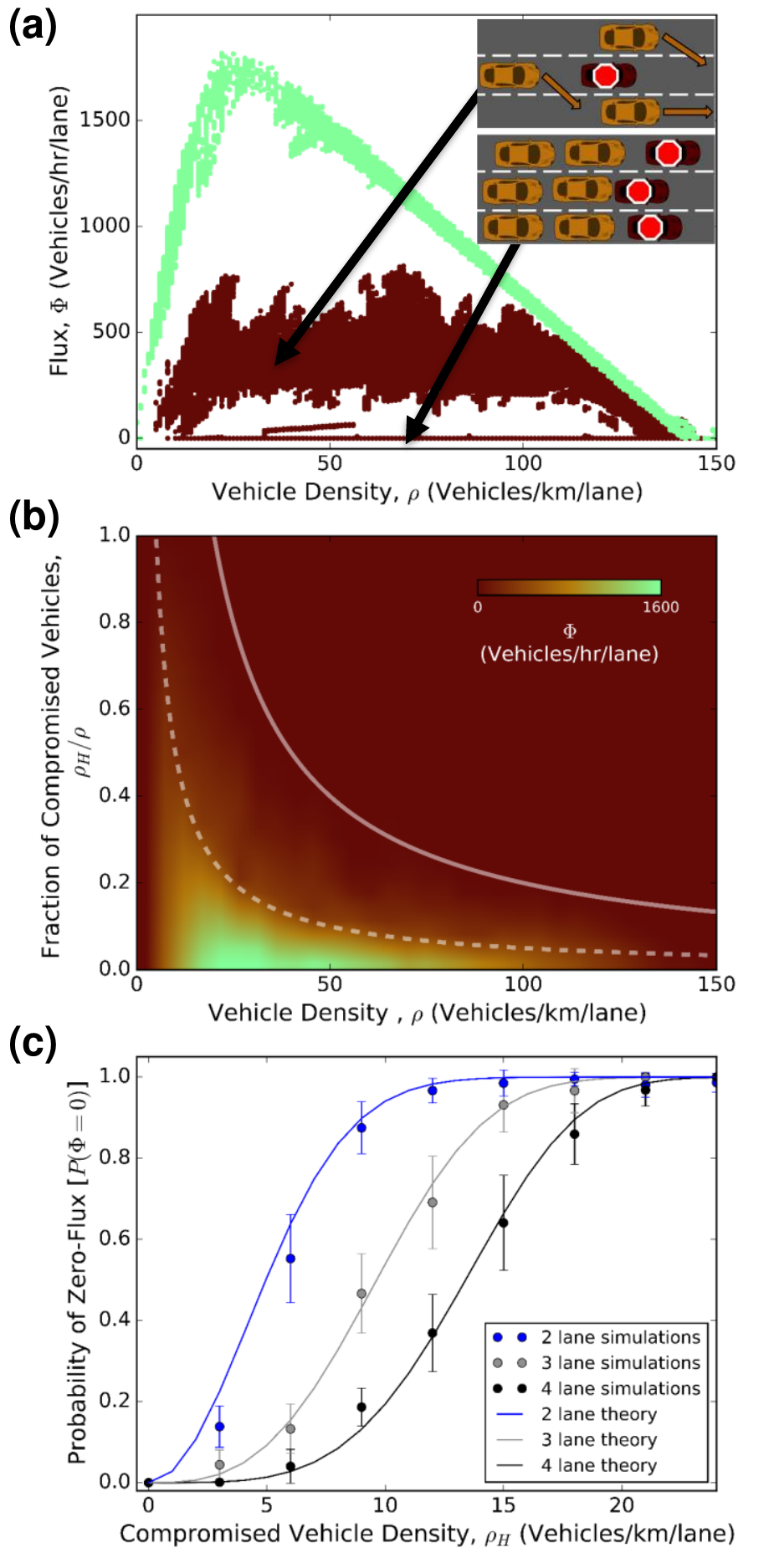}
\caption{Disruption of vehicle traffic caused by hacking on individual roads.  (\textbf{a}) Vehicle flux $\Phi$ for normal driving conditions (light green) compared to the flux after a number of vehicles are disabled by hacking (dark red).  Simulations explore $0 \le \rho \le 150$ vehicles/km/lane, and a varying fraction of disabled vehicles $0 \le (\rho_H/\rho) \le 1$ (dark red). Vehicle flux post-hack causes a bifurcation of the data with $\approx 15\%$ of the simulations having residual flow ($\Phi > 0$ vehicles/km/lane), while the remaining 85\% have no flow ($\Phi = 0$ vehicles/km/lane).  Insets schematically illustrate the traffic flow patterns.  (\textbf{b}) Data in (a) plotted as a heatmap.  Lines correspond to contours of constant compromised vehicle density with $\rho_H = 5$ (dashed) and 20 (solid) vehicles/km/lane.  (\textbf{c}) Probability that agent-based IDM/MOBIL simulations produce a zero-flux outcome (dots) compared to the predictions of percolation theory (solid lines).}
\label{fig:fig2}
\end{figure}

To begin our examination of how hacking targeted at Internet-connected vehicles disrupts traffic, we construct a minimal agent-based model for traffic flow. We simulate individual vehicles with the Intelligent Driver Model (IDM)~\cite{helbing2001,treiber2000,chen2010calibration}, a time-continuous car-following set of equations that quantitatively reproduces known traffic dynamics (Methods). To model lane changing, we use the Minimizing Overall Breaking Induced by Lane changes (MOBIL) framework~\cite{treiber2006mobil, treiber2009modeling} (Methods).  Briefly, a vehicle changes lanes if doing so would allow it and its neighbors to better match their preferred speeds. As a validation for these simulations, we varied the density of vehicles, $\rho$ (number of vehicles/km/lane), and measured the flux, $\Phi(\rho)$ (number of vehicles/hr/lane), on a 3 lane road. These simulations reproduce the classic peaked relationship between traffic flux and density (Fig.~\ref{fig:fig2}a, green symbols), which has been observationally measured and theoretically reproduced in agent-based and continuum models~\cite{helbing2001,lwr1955,nagel1992}.

A variety of hacking scenarios targeting Internet-connected vehicles lead to the same generic outcome where compromised vehicles stop and become obstacles on the road. We simulate these scenarios by randomly selecting vehicles and marking them as compromised, thus halting their motion.  We then simulated post-hack traffic on a straight 3 lane road with periodic boundary conditions to maintain constant total vehicle density. We investigated realistic densities ranging from $\rho=1$ to 150 vehicles/km/lane, and fraction of compromised vehicles ranging from 0\% to 100\% of all vehicles. Post-hack, we observe decreased flux at every density (Fig.~\ref{fig:fig2}a, dark red symbols). Interestingly, these data separate into two distinct regimes. Traffic continues to flow in $\approx$ 15\% of the simulations, albeit at a significantly reduced rate (Fig. 2a, red band centered on $\Phi$ $\approx 400$ vehicles/hr/lane).  More strikingly, the remaining $\approx$ 85\% of the simulations lead to a complete loss of traffic flow (Fig.~\ref{fig:fig2}a, red data at $\Phi = 0$ vehicles/hr/lane).  Evidently, there are two distinct phenomenologies arising post-hack: one where traffic is slowed, and another where traffic is stopped.  The gap between these two $\Phi(\rho)$ curves (Fig.~\ref{fig:fig2}a, white region where $\Phi \approx 100$ vehicles/hr/lane) is broadly independent of $\rho$, suggesting a categorical distinction between the two flow phenomena and ruling out a continuous transition between the flow and no-flow states.  

To better understand the effects of disabled vehicles on traffic flow, we plot the same simulation data as a flux heatmap with varying density and fraction of compromised vehicles (Fig.~\ref{fig:fig2}b).  Most of the heatmap's area corresponds to zero-flux outcomes (Fig.~\ref{fig:fig2}b, dark red), echoing the observation that $\approx$ 85\% of the simulations lead to a complete loss of traffic flow.  Intriguingly, contours of constant $\Phi$ coincide with contours of constant compromised vehicle density, $\rho_H$ (Fig.~\ref{fig:fig2}b, white solid and dashed lines).  

There are two classic flow phenomena that can produce zero-flux in the presence of constrictions: clogging~\cite{thomas2015, zuriguel2014,peter2017} and percolation~\cite{sahini2014}. In the first case, interactions between objects produce configurations that prevent other objects from flowing past each other, eventually arresting flow. Thus, clogging is a kinetic phenomenon, and the typical time it takes for a clog to form depends on the density of both mobile constituents and obstacles~\cite{peter2017}.  In fact, clogging is the phenomenological flow we commonly see when a vehicle breaks-down or traffic congestion increases during rush hour; in both cases, traffic flux gradually decreases over an expanding stretch of road.  In contrast, percolation occurs when a continuously connected obstruction spans the system, and is therefore a purely geometric phenomenon ~\cite{sahini2014}.  Thus, while clogging is a slow build-up to reduced traffic, percolation is a sudden and abrupt transition from flow to no-flow states independent of the density of free-flowing vehicles.  Our observations that flux contours are consistent with $\rho_H$ rather than $\rho$ (Fig.~\ref{fig:fig2}b, white solid and dashed lines) suggests that geometric percolation of compromised vehicles is the underlying zero-flux mechanism in post-hack traffic considered here (Supplementary Materials).

\subsection*{Analytical expression for post-hack traffic flow}

If post-hack traffic flow is a percolation flow phenomenon, an analytical expression for disabled vehicles to randomly align into geometric blockages should predict the probability of zero-flux traffic.  We derived this expression to account for an arbitrary number of lanes, $\ell$, the density of compromised vehicles per lane, $\rho_H$, the length of the road, $L$, and the minimum center-to-center distance between two vehicles on adjacent lanes that still allows a third vehicle to lane-change between them, $s$. That is, $s$ is twice the vehicle length, which is held constant for simplicity. Assuming vehicles are distributed uniformly at random throughout the road prior to the hack, we find the probability of percolation, $P_p$, (Methods) to be:
\begin{equation}
P_p=1-\Bigg[  1- \left( \frac{s}{L} \right)^{\ell-1}\cdot \left( 2 - \frac{s}{L} \right)^{\ell -1}   \Bigg]^{({L \, \cdot \, \rho_H}){^\ell}}.
\label{eq:eq2}
\end{equation}
This expression gives the probability that compromised vehicles are positioned in such a way as to block all lanes of a multi-lane road. While we are interested in varying $\ell$, $\rho_H$, and $L$ to account for different traffic conditions and magnitudes of hacks, the effective vehicle length is fixed at 7 m ($s=14$~m), which corresponds to the typical separation between cars in dense traffic~\cite{treiber2000}.

We simulated hacking events of different magnitudes, as measured by the number or fraction of vehicles compromised during the hack, and measured the flux to determine whether zero-flux events occur as frequently as predicted by percolation in Eq.~(\ref{eq:eq2}). Indeed, our analytical expression accurately captures the relationship between the probability of zero-flux, the density of compromised vehicles, and the number of lanes (Fig.~\ref{fig:fig2}c solid lines), with a remarkably high coefficient of determination in each case ($R^2>0.99$). This is consistent with our hypothesis that geometric percolation causes standstill traffic in the post-hack conditions we explore; if clogging played a significant role, zero-flux traffic would have occurred more often than predicted by this analytical expression (Supplementary Materials) ~\cite{zuriguel2014,thomas2015}. 

Beyond computational simulations, human drivers self-organize on roads forming spatial distributions that may differ from those formed by the combined IDM/MOBIL model.  To address this potential concern and validate our percolation formula, we made use of the NGSIM dataset~\cite{NGSIM}, which is a US Department of Transportation-funded measurement of driver spatiotemporal trajectories.  Similar to the procedure used to analyze simulations, we randomly selected a subset of vehicles to be hacked, compromised, and disabled (Supplementary Materials). We found our analytical model again captures the percolation probability with high accuracy ($R^2>0.99$), even when applied to the empirical NGSIM data.

Percolation of compromised and disabled vehicles across a road or highway represents a particularly concerning scenario, as emergency vehicles can no longer use roads that become totally blocked. Furthermore, this analysis shows that zero-flux can occur with surprisingly low densities of disabled vehicles. For example, with just 6 compromised vehicles/km/lane ($< 5\%$ of cars in bumper-to-bumper traffic) the probability of percolation across a two lane road is $\approx 60 \%$. Fortunately, as percolation is a geometric effect, the probability that a hacking event will block a road can be directly calculated with Eq.~(\ref{eq:eq2}) for any set of parameters, circumventing the need for time-consuming, model-specific, agent-based simulations. Thus, this mathematical insight enables us to rapidly assess the risk of zero-flux traffic for any road, with any number of compromised connected vehicles.

\subsection*{Compromised vehicles gridlock Manhattan}

\begin{figure*}
\begin{centering}
\includegraphics[width=\textwidth]{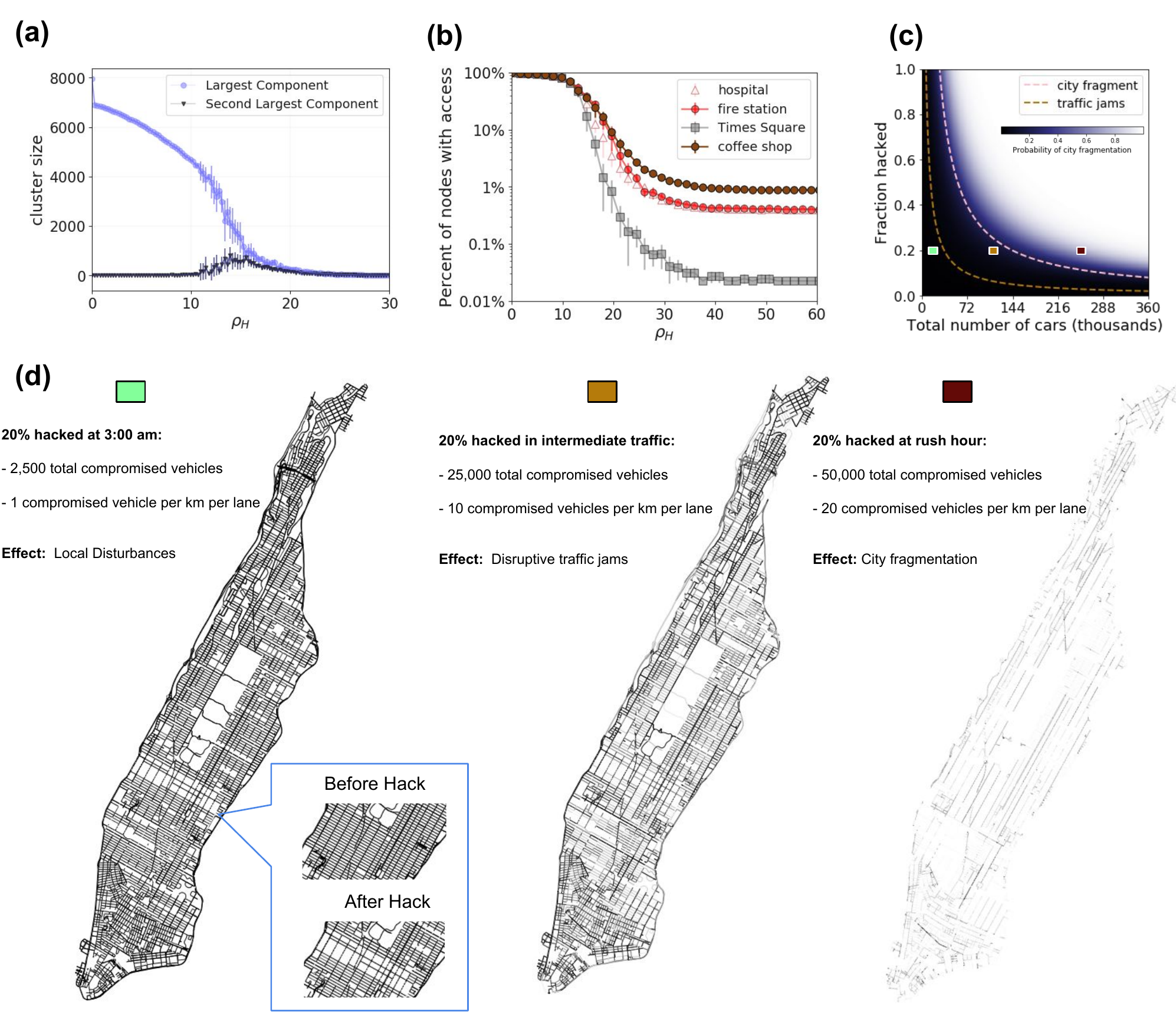}
\caption{Consequences of Internet-connected vehicles being disabled by hacking on a city street network. (\textbf{a}) Size, measured by the number of connected street intersections in the largest and second largest connected components of the street network, as a function of the density of compromised vehicles $\rho_H$. At the critical compromised vehicle density, $\rho_H \approx 13$ compromised vehicles/km/lane, the size of the second largest component reaches its maximal value and becomes comparable to the size of the largest component. This critical density represents the point at which the network begins to fragment into roughly equal-sized subnetworks, called the point of ``city fragmentation." Fluctuations in cluster size increase near the critical point, as expected for percolation-based phase transitions.  The jump in cluster size for $\rho_H$ just larger than 0 compromised vehicles/km/lane is due to all single-lane roads being blocked. (\textbf{b}) Plot of the fraction of nodes (street intersections) with access to coffee shops, emergency services, and Times Square (a example landmark of interest) as the density of compromised vehicles is varied.  (\textbf{c}) The horizontal axis is bounded by an estimate for the maximum number of vehicles that can fit bumper to bumper on all the roads of Manhattan island. Hyperbolic contours correspond to constant $\rho_H$. Pink dashed line corresponds to the critical threshold $\rho_H=13$ compromised vehicles/km/lane. Dashed orange line corresponds to $\rho_H=5$ compromised vehicles/km/lane, which is the compromised vehicle density that generally corresponds to the onset of disruptive traffic jams.  (\textbf{d}) City networks with edges (streets) shaded by the probability that they become totally blocked based on Eq.~(\ref{eq:eq2}).  Lighter shades correspond to higher probabilities of obstructions blocking traffic flow. Lower-left inset shows the local disruption that can result from even a small-scale hacking event. Green/orange/red colored rectangles above the city networks correspond to the compromised vehicle density and total number of connected vehicles shown by the identically colored rectangles in (\textbf{c}).}
\label{fig:fig3}
\end{centering}
\end{figure*}

To quantify the broader cyber-physical risks posed by hacking targeted at Internet-connected vehicles, we must investigate how traffic flow in an entire urban street network is affected. Using our percolation-based analytical formula Eq.~(\ref{eq:eq2}), we can directly and immediately compute the likelihood of any road being blocked, which saves significant computational time otherwise required for city-scale agent-based simulations. Thus, we connect road-level traffic dynamics to network-level structure by stochastically marking roads as obstructed according to their percolation probabilities. We can then use tools from network theory to assess the degree of urban disruption due to hacking. While previous studies have investigated urban street network robustness to both random and targeted pruning of edges~\cite{albert2000, callaway2000, abbar2016, wang2015}, the consequences of a hack have not been directly explored.  The critical advancement introduced by percolation, therefore, is to motivate the pruning of edges based on the underlying traffic features quantified by $\rho_H$, $L$, $\ell$, and $s$ rather than the bare network structure. 

We applied this approach to the island neighborhood of Manhattan, in New York City, USA, using map data from the Open Street Maps tool OSMnx~\cite{boeing2017osmnx}. After fixing the density of compromised vehicles, each road is stochastically set to be accessible or blocked according to the probability calculated with Eq.~(\ref{eq:eq2}), which depends on the length and number of lanes for each road (Methods). Using this method, we quantify the degree to which a hacking event disrupts the city by measuring the city's connectivity. In our analysis, a connected component represents a spatial network of roads accessible to each other but not accessible to the rest of the network. The size of each connected component is computed as being equal to the number of accessible street intersections (nodes) within that component. We compute the size of the largest and second largest connected components; when the size of the largest component becomes comparable to the size of the second largest component, the city network has been fragmented~\cite{li2015percolation,wang2015,abbar2016}. Just as percolated compromised vehicles block individual streets, stochastically blocked individual streets can percolate across the city network~\cite{shlomo2018percolation,zeng2019switch}, which can be described as a percolation-of-percolations event.

We find that for $\rho_H \lessapprox 10$ compromised vehicles/km/lane, small subnetworks are broken off of the largest connected component (Fig.~\ref{fig:fig3}a, dark purple lower line). At a critical compromised vehicle density of $\rho_H \approx 13$ compromised vehicles/km/lane, the number of nodes in the second largest connected component reaches its maximal value and is comparable to the number of nodes in the largest connected component. Thus, $\rho_H = 13$ compromised vehicles/km/lane represents the critical point at which there is no longer a substantial network of functional roads that connects points through the city. Above this compromised vehicle density, the city has fragmented. For $\rho_H \gtrapprox 20$ compromised vehicles/km/lane almost all the roads in the city are blocked, and thus we see how compromised vehicles disrupt urban traffic via the percolation-of-percolations.

Along with severe traffic gridlock, access to hospitals and fire stations will be affected in the event of a large-scale hack. To quantify how this disruption affects access to these essential services, latitude and longitude locations of these services were obtained with the Google Places API, and mapped to the closest corresponding intersection in the street network (Methods). At low $\rho_H$ (Fig.  3b,  $\rho_H  \lessapprox  10$ compromised vehicles/km/lane), nearly every service is accessible from anywhere in the city. Once city fragmentation occurs, access to services decreases dramatically. At very large $\rho_H$ (Fig.  3b,  $\rho_H  \gtrapprox  30$ compromised vehicles/km/lane), the only intersections with access to services are the intersections that \textit{contain} the services. At these large densities of hacked, compromised, and disabled vehicles, all the curves plateau to the fraction of nodes with a given service (Fig.~\ref{fig:fig3}b). For example, if $\approx 0.5\%$ of intersections contain hospitals, the hospital access curve plateaus to $0.5\%$.  We find that, despite emergency services being well distributed throughout the city (Supplementary Materials), access to these services is still dangerously diminished in the event of a large-scale hack.

We quantify the risk of city-wide disruption through the probability of network fragmentation.  This probability is defined by the density of compromised vehicles where the size of the second-largest connected component of the street network is maximized.  Our measurement of city-wide risk is based on the total number of vehicles on Manhattan roads, and the fraction that are compromised (Fig.~\ref{fig:fig3}c). We find that when either the traffic density or the fraction of compromised vehicles is very low (number of compromised vehicles $\lessapprox 2,500$), the probability that compromised vehicles percolate and block individual streets is negligible, so the city grid as a whole remains well connected ($>95\%$ of edges retained, (Fig.~\ref{fig:fig3}d, left).  Nevertheless, local disruptions can be significant due to stochastic variations. For example, one random instantiation found 2,500 compromised vehicles distributed throughout Manhattan blocked over 70\% of roads in the neighborhood just south of Central Park (Fig.~\ref{fig:fig3}d, left).  While these disruptions are not widespread enough to fragment the city, their impact increases during intermediate traffic conditions due the increased number of mobile and compromised vehicles (Fig.~\ref{fig:fig3}d, middle).  Not surprisingly, the potential for city-wide disruption peaks during rush-hour when traffic density is at its highest and the chances of individual streets being blocked exceeds 50\% (Fig.~\ref{fig:fig3}d, right).  The low, medium, and high traffic density regimes considered here (Fig.~\ref{fig:fig3}d, left to right), correspond to an average probability of blocked roads of less than 25\%, greater than 25\% but less than 50\%, and greater than 50\%, respectively (Supplementary Materials). The sensitive dependence of city-wide percolation on $\rho_H$ indicates that the overall risk rapidly increases as it approaches 13 compromised vehicles/km/lane, which ultimately leads to a cascade of consequences from the inability to access most parts of the city.

\subsection*{Large-scale hacking below the percolation threshold}

Hacking events that fall below the threshold for city fragmentation can cause significant disruption and danger.  As we have already shown, a small-scale hack can stochastically incapacitate a localized region of the city (Fig. 3d, left inset).  Alternatively, the same small-scale hack can induce the more familiar phenomenology of clogging simply by disabling a handful of Internet-connected vehicles and waiting for traffic to build up.  There are even second-order effects where {\it both} clogging {\it and} localized percolation happen {\it simultaneously}, which could potentially result in even wider disruption due to non-linear interactions between kinetic (clogging) and non-kinetic (percolation) flows.  To better grasp the potential disruption of sub-critical densities of disabled vehicles, we examined post-hack traffic dynamics on a Manhattan-like grid, using the Simulation of Urban Mobility (SUMO) \cite{sumo} traffic suite (Supplementary Materials).  In these simulations, we observed clogging-like kinetic slowing traffic at densities below the percolation threshold.  As expected, the critical threshold for $\rho_H$ at which the average vehicle velocity drops to zero coincides with the percolation threshold.  While these additional sub-percolation simulations shift away from static blockages to examine time-dependent dynamics, they underscore the reality that a percolation-of-percolation event exists within a wider ecosystem of  disruptions.

Our effort to isolate the specific nature of percolation and distinguish its statistical properties from clogging may appear to underestimate the severity of hacking targeted at Internet-connected vehicles.  However, this distinction is critically important for developing risk mitigation, response, and recovery plans.  For example, a plan that is highly effective for clogging at a sub-critical density of disabled vehicles, may be significantly less effective at larger densities of disabled vehicles, and of course, vice versa.  Furthermore, our analytical approach with Eq.~(\ref{eq:eq2}) captures the underlying risk of percolation-based gridlock, while bypassing computationally expensive large-scale traffic simulations.  This computational efficiency is appealing for developing real-time recovery plans in the aftermath of a cyber attack where case-specific details can be incorporated in the recovery response.  Indeed, with the proliferation of Vehicle-to-Vehicle (V2V) and Vehicle-to-Infrastructure (V2I) connectivity, these challenges are already permeating urban infrastructure.  New York City has current plans to install V2V and V2I technology~\cite{NY-CV,NY-CV2}, suggesting an urgent need to identify, understand, and plan ahead for the likelihood of vehicle-targeted hacking.

\section*{Conclusion}

Quantification is the critical first step in cyber-physical risk mitigation.  With the results presented here, we found just $\approx 13$ compromised vehicles/km/lane on the Manhattan street network is enough to cause citywide disruption, wherein portions of the city become disconnected from key services. This density corresponds to $\approx$ 10\% of the capacity of the city, or about 30\% of all vehicles at intermediate traffic density. From the New York State vehicle registry, the four largest vehicle manufacturers (Honda, Toyota, Ford, and General Motors) each account for around 10\% of the total number of vehicles registered (Supplementary Materials). Thus, if any one of those four manufacturers were compromised during rush hour, it would cause city-wide disruption with probability $>40\%$.  If two manufacturers were compromised, city fragmentation becomes a near certainty, occurring with probability $>95\%$. Because we have no precedent for large scale cyber-physical hacking, we cannot directly compare these predictions to empirical observations.  Indeed, records of traffic accidents across New York City show that there are at most $\sim 30$ simultaneous accidents, which is far below the percolation threshold \cite{NYCaccidents}.  Evidently, the percolation-of-percolations phenomenon described here is a flow phenomenon that is statistically unlikely to occur in conventional conditions, making a cyber-physical hack the only apparent means of observing its occurrence.

As a direct benefit of developing the percolation-of-percolation framework, we have incidentally uncovered an insight useful for developing risk-mitigating strategies.  {\it Using multiple distinct networks for connected vehicle communications and infrastructure decreases the number of vehicles that can be compromised in a single malicious cyber-intrusion.}  For example, if there were 20 compartmentalized networks in a city, each of which was responsible for 5\% of connected vehicles communications, the chance of citywide fragmentation would be low ($<10\%$) if any one of these networks was hacked.  A hacker deliberately seeking to cause a large-scale disruption faced with this compartmentalized multi-network architecture would therefore be required to execute multiple simultaneous intrusions across multiple distinct networks, increasing the cyber-attack's difficulty and makes it less likely to occur.  In conjunction with conventional cybersecurity strategies,~\cite{shamsh2013roaming,almeshekah2016} the chances of a percolation-of-percolation event could be effectively reduced to zero.  While isolated traffic disruptions are well-understood from the perspective of transportation science, the easy digital scalability and replicability of hacking means a single well-coordinated attack could surpass any familiar traffic condition.  As we explore these technology enabled ``unknown unknowns,'' we must be aware of how unintended blind spots can be exploited so that we can preemptively predict and prevent their harms.

\section*{\label{sec:Methods}Methods}

\subsection{\label{subsec:IDM}Intelligent Driver Model (IDM) Simulations}

We simulate the motion of individual vehicles using IDM. The rules for IDM simulations are:

\begin{eqnarray}
\dot{x}_n = \frac{dx_n}{dt} &=& v_n, \nonumber \\
\dot{v}_n = \frac{dv_n}{dt} &=& a \left[1-\left(\frac{v_n}{v_0} \right)^4 - \left( \frac{s^*(\Delta v_n, v_n)} {s_n}\right)\right], \nonumber \\
s^*(\Delta v_n, v_n) &=& s_0+v_nT+\frac{v_n\Delta v_n}{2\sqrt{ab}},
\end{eqnarray}
where $x_n$ and $v_n$ denote the position and velocity of the $n^{\rm th}$ vehicle, and $s_n$ is the distance between $n^{\rm th}$ vehicle and the vehicle in front of it on the same lane. 
We choose parameters~\cite{treiber2000}: $v_0 = 120$~km/h is the velocity a vehicle would drive in free traffic; $s_0 = 2$~m is the minimum acceptable gap maintained from the front bumper of one car to the rear bumper of the car in front of it in dense, standing traffic~\cite{Treiber2008}; $T = 1.6$~s is the minimum possible time for a vehicle to reach the current position of the vehicle in front of it; $a = 0.73$~m/s$^2$ is the maximum acceleration; $b = 1.67$~m/s$^2$ is the comfortable deceleration.  In simulations, we numerically solve these equations for $N$ vehicles, and after 1,000 time steps corresponding to 100 s of simulated drive time,  randomly select a portion of them to be ``compromised,'' and stop them where they are on the road.  Interestingly, the percolation results are substantially independent from the numerical values of these microscopic model parameters (Supplementary Materials).

\subsection{\label{subsec:MOBIL}MOBIL Lane Changing Rules}

To realistically model the behavior of individual vehicles, we also need a microscopic description to determine when vehicles should switch lanes.  While a variety of options have been established in the literature, here, we utilize the Minimizing Overall Breaking Induced By Lane changes (MOBIL) model\cite{treiber2006mobil,treiber2009modeling}.  This framework considers whether a vehicle and its neighbors would better match their preferred speed if the vehicle changes lanes.  Defining the change in acceleration post-lane-change between the next time step $t+1$ (after the lane change) and the current time step $t$ (before the lane change) as
\begin{eqnarray}
\Delta \ddot{x}_i & = & \ddot{x}_i(t+1) - \ddot{x}_i(t), \nonumber \\
\Delta \ddot{x}_{i-1} & = & \ddot{x}_{i-1}(t+1) - \ddot{x}_{i-1}(t), \quad {\rm and} \nonumber \\
\Delta \ddot{x}_{j-1} & = & \ddot{x}_{j-1}(t+1) - \ddot{x}_{j-1}(t),
\end{eqnarray}
we can express the MOBIL condition as
\begin{equation}
\Delta \ddot{x}_i + p(\Delta \ddot{x}_{i-1} + \Delta \ddot{x}_{j-1}) > 0,
\end{equation}
where the index $i$ corresponds to the vehicle changing lanes, $i-1$ is the current vehicle behind the lane-changing vehicle at time $t$, and $j-1$ is the vehicle that will be behind the lane changing vehicle at $t+1$ if $i$ changes lanes.  The constant $p$ is referred to as the politeness factor; we choose $p = 1$, which corresponds to force minimization of a vehicle and its nearest neighbors.

In simulations with three lanes, the left-most and right-most lanes, can change into the center lane, and at every time step, vehicles in these edge lanes initiates lane changing, only if Eq. (4) is satisfied and a random number $r_n ~\in~[0,1]<0.5$. This additional random variable $r_n$ prevents the emergence of unrealistic large-scale simultaneous lane switching. Vehicles in the center lane choose the left or right lane to change into, depending on whether $r_n<0.5$ or $>0.5$ respectively.  Within fluctuations, we find that different values for $p$ do not affect macroscopic flux-densities measurements.  Likewise, in the presence of disabled vehicles, $p$ does not change the underlying percolation-geometric transition discussed in the main text.

\subsection{\label{percolation}Percolation of Compromised Vehicles}

Given the prevalence of zero-flux events post-hack in our highway-like simulations, even when only a portion of vehicles are compromised, we sought to understand whether percolation of compromised vehicles is the dominant phenomenon after a cyber-attack. Percolation of disabled vehicles occurs when, on an $\ell$ lane road, there is an $\ell$-tuple of disabled vehicles positioned across all lanes such that no other vehicle can pass them (Fig \ref{fig:fig2}a, lower inset).

To this end, we derived an expression for the probability of a percolated configuration on an $\ell$ lane road of length $L$ and a per-lane vehicle density $\rho$, with effective vehicle size $d$. The effective vehicle size is the length from the rear bumper of one vehicle to the rear bumper of the \textit{next} vehicle in ``bumper to bumper'' traffic. Of course in bumper to bumper traffic cars' bumpers are not actually in contact, so $d$ is slightly larger than the physical length of a vehicle. In our case $d=7$~m, which is typically the separation between cars in dense traffic \cite{treiber2000}.

To determine the probability that compromised vehicles end up in a percolated position, \textit{i.e.} one that would block all lanes of a highway if the vehicles were frozen in place, we first assume that the position of the $i^{\rm th}$ connected vehicle in lane $\alpha$ will be distributed uniformly: 
 \begin{equation*}
 X_i^{\alpha}\sim U(0,L) \,\,\,  \forall i,\alpha.
 \end{equation*}

\begin{figure}
\includegraphics[width=\columnwidth]{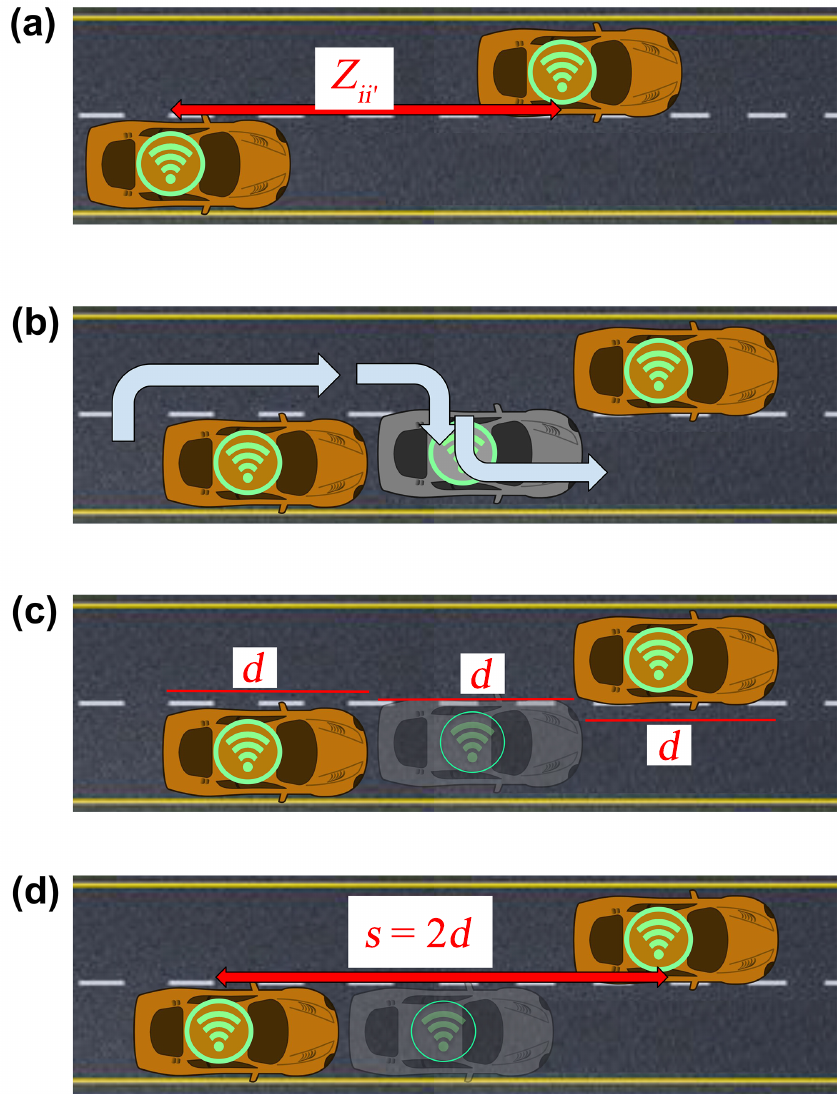}
\caption{Cartoon showing how we define $s$.  \textbf{(a)} The center-to-center distance between two vehicles $i$ and $i^{\prime}$ on lanes $\alpha$ and $\alpha+1$ is considered a random variable $Z_{i\,i^{\prime}}$. \textbf{(b)} In order for the silver vehicle to lane change around the orange vehicle in the right lane, it has to be able to fit between the two orange vehicles, following the path shown. \textbf{(c)} If the silver car can just barely fit between the orange vehicles, and each vehicle has an  effective size $d$, then \textbf{(d)} the minimum center-to-center distance between the orange vehicles that \textit{wouldn't} cause compromised vehicle percolation if the orange vehicles were suddenly hacked is $s = 2d$. Of course, while the cartoon shows three cars of actual length $d$ tightly packed bumper to bumper, in reality vehicles need a little bit of extra room to maneuver, beyond their physical size. This is why we use the effective vehicle length $d$ of $7$~m rather than the actual average length of a vehicle in calculating the minimum center-to-center distance between vehicles to avoid compromised vehicle percolation, $s$. So $s$ is the minimum required center-to-center distance between vehicles such that they are not in a position that they would cause compromised vehicle percolation upon being disabled by a hack.}
\label{fig:cartoon_for_derivation}
\end{figure}

We next define the random variable
\begin{equation*}
Z_{ii^{\prime}}=| X_i^{\alpha} - X_{i^{\prime}}^{\alpha+1}|,
\end{equation*} 
which describes the center-to-center distance between two vehicles in adjacent lanes, projected onto the direction of traffic flow. This is the quantity we are interested in, since we want to understand the probability that cars in adjacent lanes are positioned such that a third car can not lane-change between or around them. In other words, we want to know if compromised vehicles in adjacent lanes are a distance no greater than $s$ from each other, where $s$ is twice the effective car length (Fig. \ref{fig:cartoon_for_derivation}).  If $Z_{i\,i^{\prime}}\geq s$ for all pairs of cars $i$ and $i^{\prime}$ in adjacent lanes, then percolation of compromised vehicles did not occur; a configuration that would cause compromised vehicle percolation across two adjacent lanes occurs with probability $P(Z_{i\,i^{\prime}}<s)$.

 Since $Z_{i\,i^{\prime}}$ is a random variable related to the difference between two uniformly distributed random variables with known probability distribution functions, we can directly calculate $P(Z_{i\,i^{\prime}}<s)$ using convolution. In fact, we derive this probability for an arbitrary real number $z$. For example let's consider cars labeled by $i=1$ and $i^{\prime}=2$. The probability distribution functions for their positions, $X_1$ and $X_2$, are 
 
 \begin{equation*}
 f_{X_1}(x)=f_{X_2}(x)=
 	\begin{cases}
    1/L, & \rm{if\, 0\leq x \leq L},\\
    0, & \rm{otherwise}.
    \end{cases}
 \end{equation*}
 We seek the cumulative distribution function for $z$, because this will give us the probability that $Z_{i\,i^{\prime}} < z$ for any arbitrary real number $z$.  The cumulative distribution function is:
 \begin{align*}
 F_{Z_{12}}(z)&=P(Z_{12}\leq z)=P(|X_1-X_2|\leq z),\\
 &=P(-z \leq X_1 - X_2 \leq z),\\
 &=P(X_1-X_2\leq z) - P(X_1-X_2 \leq -z),\\
&= F_{X_1-X_2}(z) - F_{X_1-X_2}(-z).
\end{align*}
 By taking derivatives and using the chain rule, we can get the probability distribution functions:
 \begin{equation*}
 f_{Z_{12}}(z)=f_{X_1-X_2}(z)+f_{X_1-X_2}(-z)
 \end{equation*}
 Note that $f_{X_1-X_2}(z)$ is the convolution of $f_{X_1}(z)$ with $f_{-X_2}(z)$, and that we know $X_1$ and $X_2$ are distributed uniformly on the interval $[0,L]$.  Then we can write, compactly: 
 
 \begin{align*}
 f_{X_1-X_2}(z)&=\frac{1}{L^2}\int_0^L{ \int_0^L { \delta((x_1-x_2)-z)\,dx_1dx_2,     }    }\\
 \\
 &=\frac{ L- z +2Lz\Theta(-z)}{L^2}.
 \end{align*}
Where $\delta(\ldots)$ is the delta function and $\Theta (\ldots)$ is the Heaviside step function. This is the standard triangular distribution, sometimes called the uniform difference distribution. The integration must be performed carefully here, because the cases when $z\geq 0$ and $z<0$, and when $x_1 \geq x_2$ and $x_1 < x_2$ must each be considered separately~\cite{Blitzstein2014}. We notice right away that this expression is an even function of $z$ and so $f_{X_1-X_2}(z) = f_{X_1-X_2}(-z)$ and so $f_{Z_{12}}(z)= 2 f_{X_1-X_2}(z)$. We are only interested in distances, so without loss of generality  we focus on $z>0$.  Thus, we can simplify: 
  \begin{align*}
 f_{Z_{12}}(z)&=\frac{2}{L^2}(L-z).
 \end{align*} 
To find $P(Z_{12}<z)$ we integrate:
 \begin{equation*}
  P(Z_{12}< z)=\frac{2}{L^2}\left(Lz-\frac{1}{2}z^2\right)
  \end{equation*}
Nothing on the right hand side of the above equation depends on the vehicle labels $1$ and $2$. This observation is generally true for any pair of vehicles on adjacent lanes. In addition, this independence is true for any arbitrary real number $z$. We are interested, however, specifically in the probability that two vehicles on adjacent lanes are separated by a distance less than $s$, so we define:
  \begin{equation*}
    P_{\rm{pair\,blocks}}=P(Z_{ii^{\prime}}< s)=\frac{2}{L^2}\left(Ls-\frac{1}{2}s^2\right).
    \end{equation*}
This is the probability that a pair of vehicles in adjacent lanes is positioned such that it would \textit{completely block} both of those lanes if the vehicles were frozen in place. For a two-lane road, this would completely block the motion of vehicles. When distributing an $\ell$-tuple of vehicles on an $\ell$ lane road, the probability that the tuple does \textit{not} block the entire road is given by
 \begin{equation*}
 P_{\rm tuple\,\rm clear}=1-P_{\rm pair\,\rm blocks}^{\ell-1}
 \end{equation*} as there are $\ell-1$ pairs of lanes to consider.

Finally, we can calculate the overall probability that \textit{any} tuple \textit{does} block the highway. To do so, we consider every $\ell$-tuple of vehicles that could potentially percolate, and find: \begin{equation*} P_{p}=1-P_{\rm tuple\,\rm clear}^{n^\ell}\end{equation*} 
where $n$ is the number of vehicles in each lane, and there are $\approx n^l$ tuples that could each potentially block the highway. In this last step, we assumed $n_j \approx n_{j^{\prime}}$  $\forall j,j^{\prime}$, i.e., there are equal numbers of vehicles in each lane. This was not strictly enforced in simulations, yet simulations still agreed with the analytical formula. Collecting all of these terms, and replacing $n$ with $L \cdot \rho_H $ we find the result for Eq.~(\ref{eq:eq2}) in the main text:
 \begin{equation*}
P_p=1-\Bigg[  1- \left( \frac{s}{L} \right)^{\ell-1}\cdot \left( 2 - \frac{s}{L} \right)^{\ell -1}   \Bigg]^{({L \, \cdot \, \rho_H}){^\ell}}.
\end{equation*}

This mathematical prediction based on the hypothesis of percolation is consistent with simulations we performed varying each parameter (Supplementary Materials).  The advantage offered by this calculation over direct numerical simulations is that it allows for immediate percolation probability calculations without the need for computationally intensive numerical simulations of agent-based models.  Without this time-saving mechanism, our city level analysis would have been practically impossible.

\subsection{Intersections}
Our analysis of citywide traffic in the post-hack scenario uses Eq.~(1) to compute the probability of obstructions blocking traffic flow.  This approach neglects potential effects that occur at intersections.  We can justify this simplification with the following back-of-the-envelope calculation.  First, we define the probability, $\Delta P_p$, that a blockage occurs exactly \textit{in} an intersection of length $\ell_0$ at the end of a road segment of size $L$, as the difference in the blockage probability between the road of length $L$ and a hypothetical road of length $L +\ell_0$.  We then approximate $\Delta P_p \approx (\partial P_p / \partial L) \cdot \Delta L = (\partial P_p / \partial L) \cdot \ell_0$.  We compute the partial derivative from Eq.~(1).  Using parameters that are reasonable for Manhattan we find $\Delta P_p$ to be on the order of $10^{-4}$.  For example, if $L = 1$ km, $s = 14$ m, $\rho = 15$ vehicles/km/lane, $\ell = 2$, and $\ell_0 = 20$ m, then we find $\Delta P_p = 2 \times 10^{-4}.$  On the other hand, the total number of intersections in Manhattan is on the order of $10^4$.  Therefore, we expect that by making this simplification to ignore intersections, we are neglecting the effects of $\sim 1$ intersection within the entire city.  This back-of-the-envelope estimate provides good evidence that ignoring intersections is quantitatively reasonable for our analysis.

\subsection{\label{code}Code Availability}
The results generated and analyzed in this study were derived from custom software.  All code is available at https://github.com/dyanni3/Traffic-Simulations.

\section*{\label{sec:Acks}Acknowledgements}
SV, DY, and PJY acknowledge support from the Georgia Tech Soft Matter Incubator. JLS was independently funded.

\begin{widetext}

\section*{\label{sec:Data}Data Availability}
The datasets generated and analyzed in this study are available from the corresponding author upon request. 

\begin{itemize}
    \item Custom software is available at: \\
    https://github.com/dyanni3/Traffic-Simulations 
    \item NGSIM data is available for public access through the US Department of Transportation:  \\
    https://ops.fhwa.dot.gov/trafficanalysistools/ngsim.htm
    \item Data for vehicle speeds in Manhattan was obtained from the NYC DOT: \\
    http://www.nyc.gov/html/dot/html/about/datafeeds.shtml
\end{itemize}

\end{widetext}

\bibliographystyle{unsrt}

\end{document}